\documentclass[aps,prd,amsmath,amssymb,preprintnumbers,nofootinbib,a4paper,preprint]{revtex4-1}
\pdfoutput=1
\usepackage{amsthm}
\usepackage{graphicx}
\usepackage{color}
\usepackage{bbm}
\usepackage{pxfonts}
\usepackage{geometry}
\usepackage{epsfig}
\usepackage{multirow}

\usepackage[ margin=5pt, font=normalsize,labelfont=bf,justification=raggedright]{caption}

\definecolor{rossoCP3}{cmyk}{0,.88,.77,.40}

\begin{document}

\title{\Large  \color{rossoCP3}  Primordial Tensor Modes \\ from \\ Quantum Corrected Inflation} 
\author{Jakob Joergensen}
\email{joergensen@cp3-origins.net} 
\author{Francesco Sannino}
\email{sannino@cp3-origins.net} 
\author{Ole Svendsen}
\email{svendsen@cp3-origins.net} 

\affiliation{
\vspace{5mm} 
{ \color{rossoCP3}  \rm CP}$^{\color{rossoCP3} \bf 3}${\color{rossoCP3}\rm-Origins} \& the 
 {\color{rossoCP3} \rm Danish IAS} 
\mbox{ University of Southern Denmark, Campusvej 55, DK-5230 Odense M, Denmark}}
 \begin{abstract}
We analyze quantum corrections on the naive $\phi^4$-Inflation. These typically lead to an inflaton potential which carries a non-integer power of the field. We consider both minimal and non-minimal couplings to gravity. For the latter case we also study unitarity of inflaton-inflaton scattering.  Finally we confront these theories with the Planck and BICEP2 data. We demonstrate that the presence of nonvanishing primordial tensor modes requires sizable quantum departures from the $\phi^4$-Inflaton model for the non-minimally coupled scenario which we parametrize and quantify. We compare the results with the minimally coupled case and elucidate the main distinctive features.  \\~\\[.1cm]
{\footnotesize  \it Preprint: CP$^3$-Origins-2014-9 DNRF90 and DIAS-2014-9}
 \end{abstract}

\maketitle
 
\section{Non-minimally coupled theories with quantum potentials}
\label{nonminsec}

The underlying origin of the inflationary paradigm constitutes a prominent problem in cosmology \cite{Starobinsky:1979ty,Starobinsky:1980te,Mukhanov:1981xt,Guth:1980zm,Linde:1981mu,Albrecht:1982wi}. Inflation is traditionally modelled via the introduction of new scalar fields. Many models have been put forward to describe the dynamics of these scalar fields and their interactions with other fields, as it has been recently reviewed in \cite{Martin:2013tda}.  

On general grounds any renormalizable field theory will recieve quantum corrections to the potential. One can think of the E. Weinberg and Coleman perturbative quantum corrections to the classical scalar potential of any field theory as a simple example of these type of corrections \cite{Coleman:1973jx, Gildener}. 
We phenomenologically characterize these corrections to the $\phi^4$ theory by introducing a real parameter $\gamma$ as follows:  
\begin{align}
V_{eff} =\lambda  \phi^4  \left(\frac{\phi}{\Lambda} \right)^{4 \gamma } \ ,
\end{align}
with $\Lambda$ a given energy scale. Of course, model by model, one can compute the specific potential as in \cite{Okada:2010jf}. Nevertheless we will show that it is possible to provide useful information on a large class of models corresponding to different values of $\gamma$ using this simple approach.
For completeness we analyze the cases in which $\phi$ couples both minimally and non-minimally to gravity. We find that for the non-minimally coupled case, the recent results by BICEP2 indicating the presence of primordial tensor modes \cite{Ade:2014xna} constrains $\gamma$ to lie in the region $0.08-0.12$, at the two-sigma confidence level. However, independently on the validity of the BICEP2 results \cite{Audren:2014cea,Mortonson:2014bja}, it is fundamental to know whether quantum corrected potentials can account for nonzero tensor modes.

Interestingly, we also discover that for large primordial tensor modes the results are largely independent on the number of e-foldings. Relevant examples of non-minimally coupled models are the Higgs-Inflation model \cite{Bezrukov:2007ep} and the  ones in which the inflaton is a composite state \cite{Channuie:2011rq, Bezrukov:2011mv,Channuie:2012bv,Channuie:2013lla}.

\section{Coupling to gravity and slow-roll inflation}

We consider the action of a scalar field non-minimally coupled to gravity:

\begin{align}
\mathcal{S}_{J} =\int d^{4}x \sqrt{-g}\left[-\frac{{\cal M}^{2}+\xi\,{\phi}^2}{2}R + g^{\mu \nu} \partial_{\mu} \phi \partial_{\nu} \phi -V_{eff} \left(\phi \right) \right].  \label{jordanaction} 
\end{align}

The subscript $J$ refers to the Jordan frame, and indicates that the gravity sector is not of the Einstein Hilbert form. Generally in the Jordan frame, the scalar background contributes to the effective Planck mass: $M_P^2 = \langle {\cal M}^{2}+\xi\,{\phi}^2 \rangle$. However, $\langle \phi^2 \rangle = 0$ in the present case and we can safely identify ${\cal M}$ with $M_P$.

It is rather cumbersome to analyze inflation using this action. Instead we proceed by applying a conformal transformation, which eliminates the non-minimal coupling term:

\begin{align}
g_{\mu\nu}\rightarrow\tilde{g}_{\mu\nu}=\Omega({\phi})^2 g_{\mu\nu},\quad\Omega({\phi})^2=1+\frac{\xi\phi^2}{M_{p}^2}.
\end{align}

We then land in the Einstein frame, in which the gravity sector is of the Einstein Hilbert form (tildes are omitted to ease notation):

\begin{align}
\mathcal{S}_{E} =\int d^{4}x \sqrt{-g}\left[ -\frac{1}{2} M_{p}^2 \,\, R+ \Omega^{-2} \left(1 +  3 M_p ^2 {\Omega'} ^{2} \right)g^{\mu \nu} \partial_{\mu} \phi \partial_{\nu} \phi  - \Omega ^{-4} V({\phi}) \right].
\end{align}

The transformation leads to an involved kinetic term and it is convenient to replace $\phi$ by a canonically normalized field $\chi$ using a field redefinition:

\begin{align}
\frac{1}{2} \left( \frac{d \chi}{d \phi} \right)^2 &= \Omega^{-2} \left(1 + 3 M_p ^2 {\Omega'} ^2 \right) = \frac{M_p^2 \left(M_p^2 + \left(1+3 \xi \right) \xi \phi^2 \right)}{\left(M_p^2+\xi \phi^2 \right)^2}. \label{1defchi}
\end{align}

The Einstein frame action then describes a scalar field minimally coupled to gravity:

\begin{align}
\mathcal{S}_{E} &=\int d^{4}x \sqrt{-g}\left[-\frac{1}{2} M_{p}^2 g^{\mu \nu}R_{\mu \nu} + \frac{1}{2} g^{\mu \nu} \partial_{\mu} \chi \partial_{\nu} \chi- U(\chi)  \right], \quad U(\chi) \equiv \left( \Omega^{-4}V_{eff} \right) \left( \phi \left( \chi \right) \right). \label{einsteinframeaction} 
\end{align}

We will assume that inflation takes place in the large field regime $\phi \gg \frac{M_p}{\sqrt{\xi}}$. In this limit the solution to \eqref{1defchi} is:

\begin{align}
\chi \simeq \kappa M_p \ln \left( \frac{\sqrt{\xi} \phi}{M_p} \right), \quad \kappa \equiv \sqrt{\frac{2}{\xi}+6 }. \label{defchi}
\end{align}

In the large field limit the Einstein frame potential then takes the form:

\begin{align}
U \left( \chi\right) = \Omega^{-4} V \left(\phi \left( \chi\right) \right) &= \frac{M_p^4}{\left(M_p^2 + \xi \phi^2 \right)^2} \lambda \phi^4 \left(\frac{\phi}{\Lambda} \right)^{4 \gamma }  \\
&= \underbrace{ \frac{\lambda M_p^4}{\xi^2}\left(1+ \exp \left[\frac{ - 2 \chi}{\kappa M_p }\right] \right)^{-2}}_{\phi^4\text{-Inflation}} \underbrace{ \left(\frac{M_p}{\sqrt{\xi}\Lambda} \right)^{4\gamma} \exp \left[ \frac{ 4 \gamma \chi}{\kappa  M_p}\right]}_{\text{Corrections from}\,\, \gamma}.
\end{align}

The underbraced '$\phi^4$-Inflation'-term refers to the potential one would obtain by setting $\gamma=0$, that is, non-minimally coupled $\phi^4$-Inflation. Large field asymptotic flatness of this term is what makes Higgs-Inflation viable \cite{Bezrukov:2007ep}. However, quantum corrections which we parametrize by $\gamma$, may spoil this feature of the potential. 

The analysis of inflation in the Einstein frame is straightforward. We proceed by the standard slow-roll approach and compute the slow-roll parameters in the large field limit using the field $\chi$ and its potential $U \left(\chi \right)$. These may be expressed in terms of the Jordan frame field $\phi$ by reinserting \eqref{defchi}:

\begin{align}
\epsilon = \frac{M_{p}^2}{2} \left( \frac{dU / d \chi}{U} \right)^2 \sim \underbrace{ \frac{8 M_p^4}{\kappa^2 \xi^2 \phi^4}}_{\phi^4-\text{Inflation}} +\frac{16M_p^2 }{\kappa^2 \xi \phi^2} \gamma+ \frac{8 }{\kappa^2}\gamma^2 . \label{epsexpanded}
\end{align}

\begin{align}
\eta = M_{p}^2 \left( \frac{d^2U / d \chi^2}{U} \right) \sim \frac{8}{\kappa^2} \left( \underbrace{- \frac{M_p^2}{\xi \phi^2}+ \frac{3M_p^4}{\xi^2 \phi^4}}_{\phi^4-\text{Inflation}}+ \frac{4 M_p^2}{\xi \phi^2} \gamma + 2 \gamma^2 \right).
\end{align}
So far $\gamma$ can assume any value and the only approximation made is the one in \eqref{defchi}. 


Inflation ends when the slow-roll approximation is violated, in the present case this occurs for $\epsilon \left(\phi_{end} \right) =1$. Thus the field value at the end of inflation is:
\begin{align}
\phi_{end} = \frac{2 M_p}{\sqrt{\xi}} \frac{1}{\sqrt{\sqrt{2} \kappa -4 \gamma}} = \left(1.07 + 0.32 \gamma \right) \frac{M_p}{\sqrt{\xi}} + \mathcal{O} \left( \gamma^2 \right) \quad \text{for} \,\, \xi \gg 1. \label{higgsresult}
\end{align}
From the first identity we derive the universal bound: 
\begin{equation}
\gamma < \frac{\sqrt{3}}{2} \ .
\end{equation}
Assuming the quantum corrections to be perturbative, in the underlying inflaton theory, we can expand for small values of $\gamma$ and obtain the right-hand side of \eqref{higgsresult}.  We set $\xi \gg 1$ since $\xi \sim 10^{4}$ is required to generate the proper amplitude of density perturbations. This is a general feature of non-minimally coupled theories of single-field inflation \cite{Channuie:2012bv, Bezrukov:2011mv, Channuie:2011rq, Bezrukov:2007ep, Lee:2014spa, Cook:2014dga}. A relatively small $\xi$ can be realized but it requires an extremely small $\lambda$ as noted in \cite{Hamada:2014iga}. We will quantify this relation between $\xi$ and $\lambda$ later, see equation \eqref{xivgamma}.

The observed Cosmic Microwave Background (CMB) modes cross the horizon about $N=60$ e-foldings before the end of inflation. The corresponding value of the inflaton field is denoted by $\chi_*$ and is given by:

\begin{align}
N = \frac{1}{M_{p}^2} \int _{\chi_{end}} ^{\chi_{*}} \frac{U}{dU /d\chi} d \chi=\frac{\kappa^2}{4}\int ^{\phi_{*}}_{\phi_{end}} \frac{1+ \frac{\xi\phi^2}{M_p^2}}{1+\gamma \left(1 + \frac{\xi \phi^2}{M_p^2} \right)} \frac{1}{\phi} d \phi
\sim \frac{\kappa^2}{8 \gamma} \ln \left[1+ \gamma \frac{\xi \phi^2}{M_p^2} \right]^{\phi_{*}}_{\phi_{end}}.
\end{align}
Combining the previous equation with \eqref{higgsresult} we deduce
\begin{align}
\phi_* & \sim \sqrt{\frac{1}{\gamma} \left( \exp \frac{8 \gamma N }{\kappa^2}-1 \right)} \frac{M_p}{\sqrt{\xi}}  \label{phiin1} \\
& =   \left( 2.83 + 5.66 \left( \frac{ \gamma N}{\kappa^2} \right) + 9.43 \left( \frac{\gamma N}{\kappa^2} \right)^2 + \mathcal{O} \left( \frac{\gamma N}{\kappa^2} \right)^3 \right) \sqrt{\frac{N}{\kappa^2}} \frac{M_p}{\sqrt{\xi}} \\
&= \left( \underbrace{8.94}_{\phi^4-\text{Inflation}} +179 \gamma + 2980 \gamma^2 + \mathcal{O} \left(\gamma^3 \right) \right) \frac{M_p}{\sqrt{\xi}} \quad \text{for} \,\, \xi \gg 1, \,\, N=60.  \label{phiin2}
\end{align}
We expanded in $\gamma$ to clarify how the result deviates from $\phi^4$-Inflation. It is evident that the $\gamma$-correction push inflation to higher field values. An expansion is, however, justified for tiny values of $\gamma$.

\section{Unitarity test via Inflaton-Inflaton scattering} 
Next, we turn to the constraints set by tree-level unitarity of inflaton-inflaton scattering. We consider the Einstein frame action in the large field regime:

\begin{align}
\mathcal{S}_{E} =\int d^{4}x \sqrt{-g}\left[ -\frac{1}{2} M_{p}^2 \,\, g^{\mu \nu}R_{\mu \nu} +\frac{M_p^2}{\phi^2} \kappa^2g^{\mu \nu} \partial_{\mu} \phi \partial_{\nu} \phi -\frac{M_p^4}{\xi^2}\lambda \left(\frac{\phi}{\Lambda} \right)^{4 \gamma} \right].
\end{align}

Violation of tree-level unitarity of the scattering amplitude, concerns fluctuations of the inflaton around its classical homogeneous background:

\begin{align}
\phi \left(\vec{x},t \right) = \phi_c \left(\vec{x},t\right) + \delta \phi\left(\vec{x},t \right). 
\end{align}

In first approximation we neglect the time dependence of the background during the inflationary period and write $\phi_c \left ( t \right) = \phi_c$. To estimate the cutoff we expand the kinetic and potential term around the background. The kinetic term for the fluctuations then takes the form
\begin{align}
\frac{M_p^2 \kappa^2}{\phi_c^2 \left(1 + \frac{\delta \phi}{\phi_c} \right)^2} \kappa^2 \left(\partial \delta \phi \right)^2 = \frac{M_p^2 \kappa^2}{ \phi_c^2 }   \left(\partial \delta \phi \right)^2 \sum_{n=0}^{\infty} \left(n+1\right) \left(\frac{-\delta \phi}{\phi_c}\right)^n. 
\end{align}
The first term of the series, i.e. the kinetic term for a free field, may be canonically normalized by a field redefinition
\begin{align}
\frac{\delta \phi}{\phi_c} = \frac{\delta \tilde{\phi}}{\sqrt{2}\kappa M_p}.
\end{align}
The kinetic term then takes the form
\begin{align}
\frac{1}{2} \left(\partial \delta \tilde{\phi} \right)^2 \sum_{n=0}^{\infty} \left(n+1 \right) \left( \frac{- \delta \tilde{\phi}}{\sqrt{2}\kappa M_p} \right)^n.
\end{align}
Expanding the potential, the leading higher order operators take on the same form
\begin{align}
\frac{ \gamma \lambda M_p^4}{\xi^2} \left( \frac{\phi_c}{\Lambda} \right)^{4 \gamma}  \left( \frac{  \delta \tilde{\phi}}{\sqrt{2}\kappa M_p} \right)^n.
\end{align}
From these expression, we determine the cutoff of the theory which controls the physical suppression of higher order operators:
\begin{align}
\Lambda_{UC} \sim \sqrt{2} \kappa M_p.
\end{align}
This implies that the theory is valid, from the unitarity point of view, till the Planck scale.

\section{Phenomenological constraints after BICEP2}
We are now equipped to confront the inflationary potential with experiments. We start by considering the constraints set by the observed amplitude of density perturbation $A_s$~\cite{Ade:2013uln}. To generate the proper value of $A_s$, the potential should satisfy at horizon crossing $\phi_*$
\begin{align}
A_s = \frac{1}{24\pi^2 M_p^4} \left\vert \frac{U^*}{\epsilon^*}\right\vert = 2.2\cdot 10^{-9} \Leftrightarrow  \left\vert \frac{U^*}{\epsilon^*}\right\vert = \left( 0.0269 M_p \right)^4. \label{ampldens}
\end{align}

\begin{figure}[b!]
\centering
{\includegraphics[width=10cm,
height=10cm,keepaspectratio]{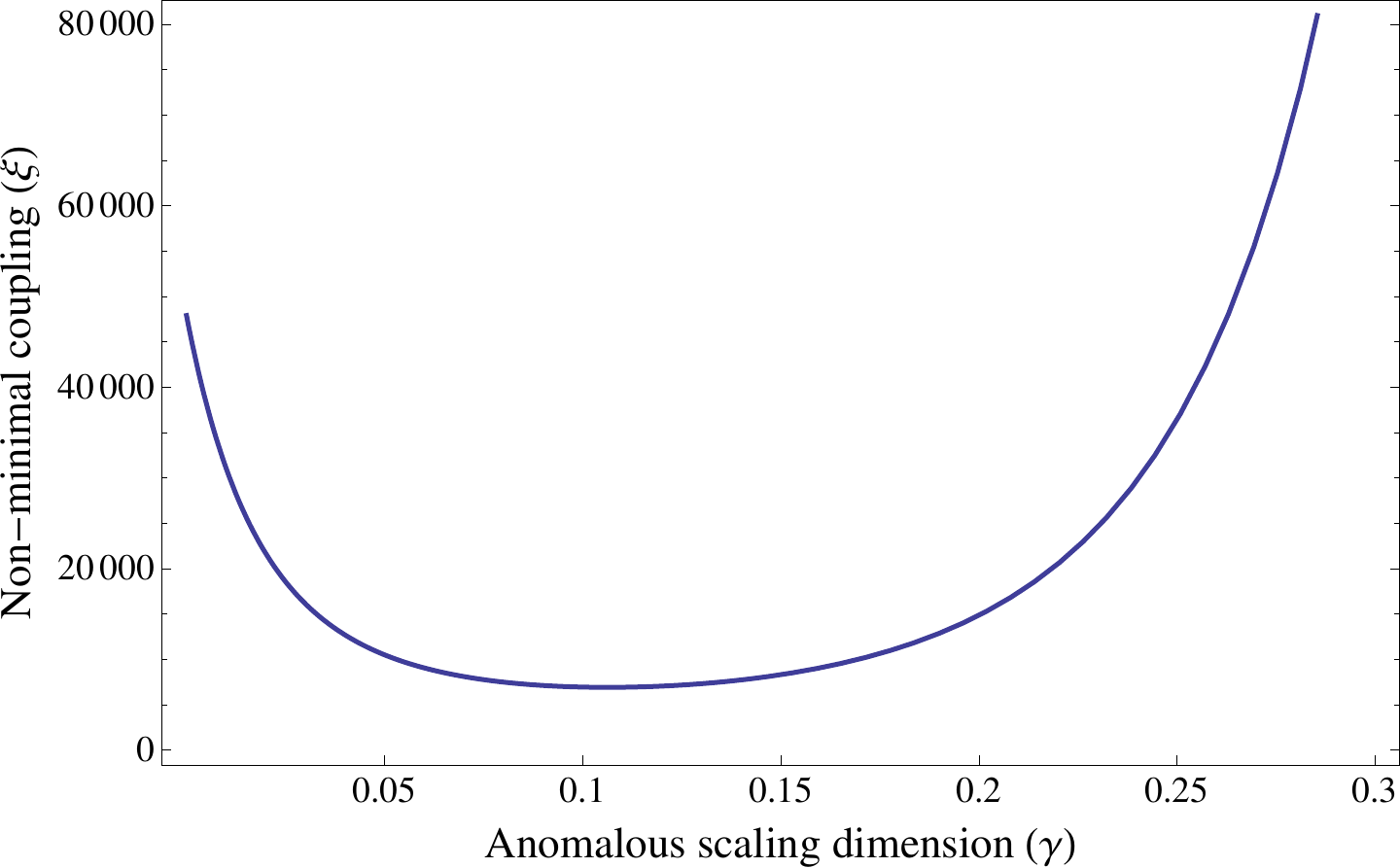}}
\caption{  Here we show \eqref{xivgamma} as a function of $\gamma$ for  $\xi \gg1$, $N=60$, $\frac{M_p}{\Lambda} =1$ and $\lambda = \frac{1}{4}$. As $\gamma$ increases from zero the magnitude of  $\xi$ needed to produce the correct amount of scalar perturbations decreases. The minimum is obtained at $\gamma \sim 0.1$. For this value of $\gamma$, the model produce an amount of tensor modes which is in agreement with the BICEP2 results, see figure \ref{rvsns}. }
\label{xivsg}
\end{figure}

For a minimally coupled quartic potential this imposes a constraint on the self coupling, which must be unnaturally small: $\lambda \sim 10^{-13}$~\cite{Guth:1982ec}. However, in the present case, \eqref{ampldens} yields a relation between $\xi$, $\lambda$ and $\gamma$. We can self-consistently solve for $\xi \gg 1$
\begin{align}
\xi = \left( \frac{3 \lambda}{4 \cdot 0.0269^4} \left(\frac{M_p}{\Lambda} \right)^{4 \gamma} \frac{\left(\exp \frac{4 \gamma N}{3}-1 \right)^{2}\left(\frac{1}{\gamma}\exp \frac{4 \gamma N}{3}- \frac{1}{\gamma} \right)^{2\gamma}}{\gamma^2 \left(\exp \frac{4 \gamma N}{3} + \gamma \right)^2} \right)^{\frac{1}{2+2 \gamma}}. \label{xivgamma}
\end{align}
The resulting constraint is plotted in Fig.~\ref{xivsg}. The magnitude of $\xi$ needed to produce the observed amplitude of scalar perturbations decreases for increasing $\gamma$ to a certain point from which it increases monotonically.

Expanding in $\gamma$ and setting $N=60$, $\lambda = \frac{1}{4}$ (which are standard values) the relation takes on a more readable form:
\begin{align}
\xi =  \underbrace{48000}_{\phi^4-\text{Inflation}} &+ (-2.27 \cdot 10^6+9.57 \cdot 10^4 \ln \frac{M_p}{\Lambda})\gamma \nonumber \\ 
&+ \left(7.46 \cdot 10^7 - 4.63 \cdot 10^6 \ln \frac{M_p}{\Lambda} +9.57 \cdot 10^4 \ln \left(\frac{M_p}{\Lambda} \right)^2 \right) \gamma^2 + \mathcal{O} \left(\gamma^3 \right). \nonumber
\end{align}

Next we consider the scalar spectral index $n_s$ and the tensor-to-scalar power ratio $r$
\begin{align}
&r=16 \epsilon_* = \frac{16}{\kappa^2} \left(  \underbrace{ \frac{8 M_p^4}{ \xi^2 \phi_*^4}}_{\phi^4-\text{Inflation}} +\frac{16M_p^2 }{ \xi \phi_*^2} \gamma + 8 \gamma^2 \right), \label{rnl}\\
&n_s = 2 \eta_*-6\epsilon_*+1  = \underbrace{1- \frac{16 M_p^2}{\xi \kappa^2 \phi_*^2} }_{\phi^4-\text{Inflation}} -\frac{ 32 M_p^2}{\xi \kappa^2 \phi_*^2} \gamma  - \frac{16 \gamma^2}{\kappa^2}. \label{nsnl}
\end{align}

\begin{figure}[b!]
\centering
{\includegraphics[width=9.5cm]{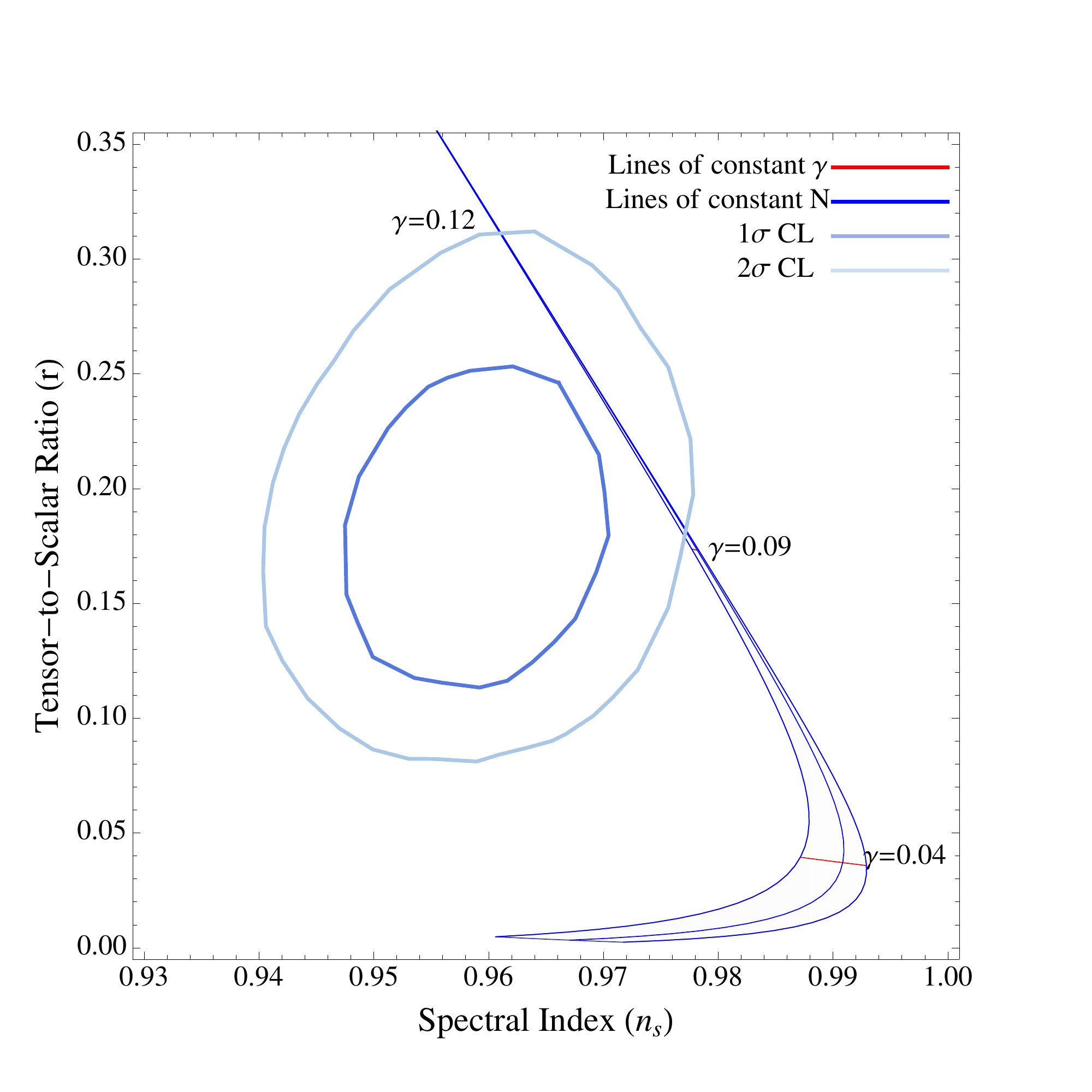}}
\caption{The cosmological parameters $r$ and $n_s$, as measured by Planck \cite{Ade:2013uln} and BICEP2 \cite{Ade:2014xna} as well as the parameters stemming from the model of this paper. The two variables $N$ and $\gamma$, span the intervals 50 to 70 and 0 to 0.15 respectively. The contours for $N$ from left to right is 50, 60 and 70.}
\label{rvsns}
\end{figure}

Using \eqref{phiin1} and expanding in $\gamma$ we obtain 
\begin{align}
r &= \underbrace{\frac{11.8}{N^2}}_{\phi^4-\text{Inflation}}+ \frac{16.3 \gamma}{N}+8.73 \gamma^2+ \mathcal{O} \left( \gamma^3 \right), \quad \text{for} \,\, \xi \gg 1 \\
&= \underbrace{0.0033}_{\phi^4-\text{Inflation}} + 0.27 \gamma + 8.73 \gamma^2 +  \mathcal{O} \left(\gamma^3 \right) \quad \text{for} \,\, \xi \gg 1, \,\, N=60.
\end{align}

\begin{align}
n_s &=  \underbrace{1-\frac{1.98}{N}}_{\phi^4-\text{Inflation}}+ \left( 1.30-\frac{3.96}{N} \right) \gamma+ \left(-0.0699-0.262 N \right) \gamma^2 + \mathcal{O} \left( \gamma^3 \right) \quad \text{for} \,\, \xi \gg 1  \\
&= \underbrace{ 0.967}_{\phi^4-\text{Inflation}} +1.23 \gamma -15.8 \gamma^2 +  \mathcal{O} \left(\gamma^3 \right) \quad \text{for} \,\, \xi \gg 1 , \,\, N=60.
\end{align}
The expansion above shows immediately that for Higgs-inflation like models\cite{Bezrukov:2007ep}, featuring a small $\gamma$, it is not possible to achieve values of $r$ consistent with the new BICEP2 results \cite{Ade:2014xna}. However, the situation changes if we allow for large corrections to the conformal $\phi^4$ potential. We use the full dependence  on $\gamma$, derived in  \eqref{phiin1}, \eqref{rnl}, \eqref{nsnl}, to plot $r$ versus $n_s$  and compare with the BICEP2 results. The comparison is in Fig.~\ref{rvsns}. We deduce that, quite {\it independently} on the number of e-foldings $N$, the two-sigma allowed value of $\gamma$ lie in the range $0.08<\gamma < 0.12$. 

\begin{figure}[t!]
\centering
\includegraphics[width=9cm]{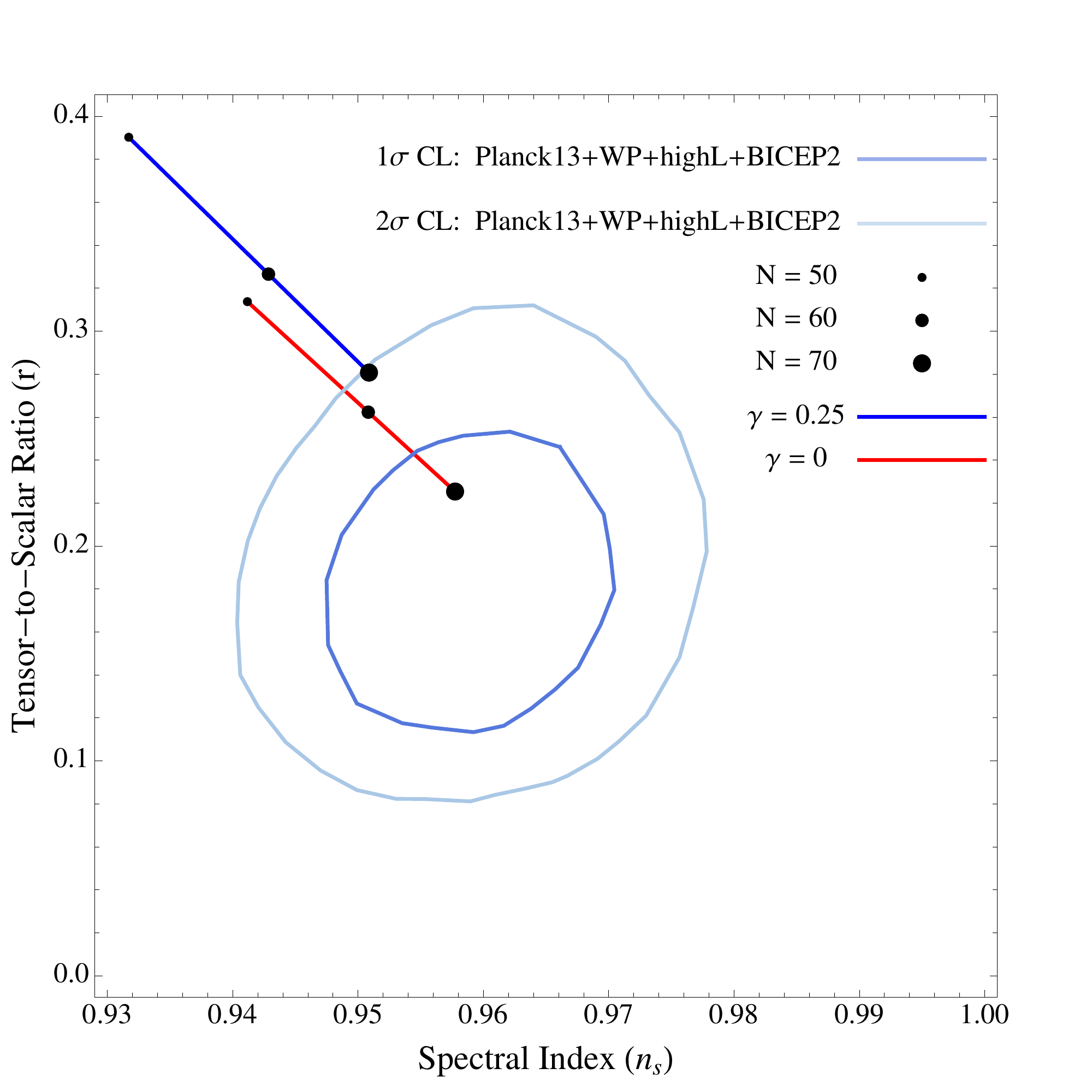}
\caption{Predictions for the minimally coupled case, given for different values of $\gamma$ and $N$. The standard $\phi^4$-Inflation is obtained for $\gamma=0$.}
\label{fig:pararvsns2}
\end{figure}
For reference we summarize the results one would obtain if the model were minimally coupled to gravity. $V_{eff}$ then produce standard minimally coupled power-law inflation. Within the slow-roll approximation, the scalar spectral index and the tensor-to-scalar ratio are:
\begin{align}
&r=16 \epsilon_* = \frac{128 M_p^2 \left(1+\gamma\right)}{\phi_i^2} = \frac{16 \left(1+\gamma \right)}{N+1+\gamma}. \\
&n_s = 2 \eta_*-6\epsilon_*+1 = 1- \frac{8M_p^2\left(1+\gamma\right) \left(3+2\gamma \right)}{\phi_i^2} = \frac{N-2-\gamma}{N+1+\gamma} .
\end{align}

These expressions corresponds to lines in the $\left(r,n_s\right)$ plane. Some of these are plotted in Fig.~\ref{fig:pararvsns2} along with constraints from Planck and BICEP2. From the figure we see that this model is consistent with the data, provided that a large fine-tuning of the self-coupling is accepted to address the amplitude of density perturbations. Contrary to the non-minimally coupled case, we observer a strong dependence on $N$ for any value of $\gamma$. Specifically for $\gamma > 0.25$ the number of e-foldings must exceed 70 to be within the two-sigma confidence level. 

To summarize  we have shown that, for the non-minimally coupled case, quantum corrections to $\phi^4$-Inflation are needed to accommodate simultaneously the latest Planck and BICEP2 results. More generally quantum corrected potentials non-minimally coupled to gravity can account for large primordial tensor modes. This result can be tested once the experimental situation is settled. 

Our analysis is sufficiently general to provide useful constraints for a general class of quantum field theories that can be used to drive inflation.


\begin{thebibliography}{99}

  
  \bibitem{Starobinsky:1979ty} 
  A.~A.~Starobinsky,
  JETP Lett.\  {\bf 30}, 682 (1979)
  [Pisma Zh.\ Eksp.\ Teor.\ Fiz.\  {\bf 30}, 719 (1979)].
  
  \bibitem{Starobinsky:1980te} 
  A.~A.~Starobinsky,
  Phys.\ Lett.\ B {\bf 91}, 99 (1980).
  
  \bibitem{Mukhanov:1981xt} 
  V.~F.~Mukhanov and G.~V.~Chibisov,
  JETP Lett.\  {\bf 33}, 532 (1981)
  [Pisma Zh.\ Eksp.\ Teor.\ Fiz.\  {\bf 33}, 549 (1981)].

  \bibitem{Guth:1980zm} 
  A.~H.~Guth,
  Phys.\ Rev.\ D {\bf 23}, 347 (1981).
  
  \bibitem{Linde:1981mu} 
  A.~D.~Linde,
  Phys.\ Lett.\ B {\bf 108}, 389 (1982).
  
  
  \bibitem{Albrecht:1982wi} 
  A.~Albrecht and P.~J.~Steinhardt,
  Phys.\ Rev.\ Lett.\  {\bf 48}, 1220 (1982).
  
  \bibitem{Martin:2013tda} 
  J.~Martin, C.~Ringeval and V.~Vennin,
  arXiv:1303.3787 [astro-ph.CO].
  
  \bibitem{Coleman:1973jx} 
  S.~R.~Coleman and E.~J.~Weinberg,
  Phys.\ Rev.\ D {\bf 7}, 1888 (1973).
  

  \bibitem{Gildener} 
  E.~Gildener and S.~Weinberg,
  Phys.\ Rev.\ D {\bf 13}, 3333 (1976).

  
\bibitem{Okada:2010jf} 
  N.~Okada, M.~U.~Rehman and Q.~Shafi,
  Phys.\ Rev.\ D {\bf 82}, 043502 (2010)
  [arXiv:1005.5161 [hep-ph]].


\bibitem{Ade:2014xna} 
  P.~A.~R.~Ade {\it et al.}  [BICEP2 Collaboration],
  arXiv:1403.3985 [astro-ph.CO].
  
  
\bibitem{Audren:2014cea}
  B.~Audren, D.~G.~Figueroa and T.~Tram,
  arXiv:1405.1390 [astro-ph.CO].

\bibitem{Mortonson:2014bja}
  M.~J.~Mortonson and U.~Seljak,
  arXiv:1405.5857 [astro-ph.CO].
  
\bibitem{Bezrukov:2007ep}
  F.~L.~Bezrukov and M.~Shaposhnikov,
  Phys.\ Lett.\ B {\bf 659} (2008) 703
  [arXiv:0710.3755 [hep-th]].

\bibitem{Channuie:2013lla}
  P.~Channuie and K.~Karwan,
  arXiv:1307.2880.

\bibitem{Channuie:2012bv}   P.~Channuie, J.~J.~Jorgensen and F.~Sannino,   
  Phys.\ Rev.\ D {\bf 86} (2012) 125035
  [arXiv:1209.6362 [hep-ph]].
  




\bibitem{Bezrukov:2011mv}   F.~Bezrukov, P.~Channuie, J.~J.~Joergensen and F.~Sannino,   
  Phys.\ Rev.\ D {\bf 86} (2012) 063513
  [arXiv:1112.4054 [hep-ph]].

\bibitem{Channuie:2011rq}   P.~Channuie, J.~J.~Joergensen and F.~Sannino,   
  JCAP {\bf 1105} (2011) 007
  [arXiv:1102.2898 [hep-ph]].


\bibitem{Lee:2014spa}   H.~M.~Lee,   
  arXiv:1403.5602 [hep-ph].

\bibitem{Cook:2014dga}   J.~L.~Cook, L.~M.~Krauss, A.~J.~Long and S.~Sabharwal,   
  arXiv:1403.4971 [astro-ph.CO].
 
  
    
%
   
   
\bibitem{Hamada:2014iga}   Y.~Hamada, H.~Kawai, K.~-y.~Oda and S.~C.~Park,   
  arXiv:1403.5043 [hep-ph].

  \bibitem{Ade:2013uln} 
  P.~A.~R.~Ade {\it et al.}  [Planck Collaboration],
  arXiv:1303.5082 [astro-ph.CO].
  
   
\bibitem{Guth:1982ec}
  A.~H.~Guth and S.~Y.~Pi,
  Phys.\ Rev.\ Lett.\  {\bf 49} (1982) 1110.
  
  
\end{thebibliography}
\end{document}